\documentclass[aps,twocolumn,showkeys,preprintnumbers]{revtex4-1}

\usepackage{graphicx}
\usepackage{dcolumn}
\usepackage{bm}
\usepackage{booktabs}
\usepackage[version=3]{mhchem}

\begin{document}


\title{Magnetic properties of the intermetallic compound \ce{HoCuBi}}
\author{C. Aguilar-Maldonado, R. Escudero}
\address{Instituto de Investigaciones en Materiales, Universidad Nacional Aut\'{o}noma de M\'{e}xico,
Apartado Postal 70-360,  M\'{e}xico D.F. 04510, M\'{E}XICO}

\email[Corresponding authors.  CA email address {cintlia@iim.unam.mx} RE email address] { escu@unam.mx}
\date{\today}

\begin{abstract}

We present an investigation on the intermetallic compound formed by  \ce{HoCuBi} which has a monoclinic crystalline $P2$ structure with lattice parameters;  $a=9.8012(26) \, \AA$, $b=6.0647(6) \, \AA$, $c=6.1663(13) \, \AA$. In this report we performed  an extensive analysis of the magnetic characteristics, specific heat at low temperature mainly close to the region of the observed antiferromagnetic transition with N\`{e}el temperature about  $7$ $K$, and to  the field induced spin-crossover; the  metamagnetic transition  typically observed in antiferromagnetic systems by measurements of  M-H, (Magnetization - Magnetic Field) at low temperatures.

\end{abstract}

\maketitle

\section{Introduction}
Many intermetallic compound of the type RT(Bi,Sb) have been   widely studied  in the last  twenty  years. The importance of these interesting compounds  has been related to the many relevant and diverse electronic properties  and many  different  characteristics  that they present \cite{Gupta2015562}. The different atomic species that can be introduced may  modified the crystalline structure, and  give place to a  number  of characteristics \cite{Szytula:1999}.  As mentioned those changes in electronic  characteristics are  depending on the   elements  forming the compound \cite{Merlo1991329} \cite{MERLO1993145} \cite{Merlo1996289}. If we consider for instances  compounds with elements type as   RT(Bi,Sb, Se), where R may be  a lanthanide atom, T  a transition element have been studied in the past. The many different   phenomena that can be present are;   Kondo effect,  and/or heavy Fermion behavior at low temperature, antiferromagnetic,  ferromagnetic characteristics, and spin glass, and also superconductivity.

The intermetallic alloys  with  rare earths atoms present  physical properties mainly related to its magnetic properties. Those alloys with  $f$-electron shells can form   different crystallographic structures and  consequently different physical  properties. In this study our  investigations were  concentrated  into the   magnetic and  thermal  behavior. The particular compound studied was HoCuBi. It presents  characteristic of spin crossover or as also named metamagnetism at low temperature,  antiferromagnetism, and the influence of the magnetic field intensity in the specific heat transition.

As above mentioned,  this work is  focused  on HoCuBi intermetallic alloy.  In the literature already published we  found no  published  study on this compound.  In the  similar  {EuCuBi} \cite{Christa:1981aa} \cite{Villars2016:sm_isp_sd_1713850} and \ce{YbCuBi} \cite{V.:2006aa},  both show  antiferromagnetic ordering with a magnetic phase transition temperature at 18 K,  and effective magnetic moment $\mu_{eff}=7.65$ BM. Also, \ce{Ho5Cu_{0.7}Bi_{2.3}} is reported \cite{MOROZKIN2005L9} with an orthorhombic $Pnma$ structure and no magnetic behavior was  reported \cite{MOROZKIN2011302}.  This is the first time a complete study was performed on this compound. 

\section{Experimental details}

The  intermetallic compound  was   prepared by using an arc melting furnace,  For the preparation   stoichiometry amounts of materials were powdered and  pastilled. Elements used were of  high purity \ce{Ho}($99.99\%$),   \ce{Cu}($99.99\%$),  and \ce{Bi} ($99.99\%$). 
The preparation was performed in the arc system with an oxygen free atmosphere in order to prevent oxygen contamination we used a Zirconium  getter and ultra-high purity argon atmosphere. Additionally,  in order to compensate bismuth losses  an excess of bismuth,  about  $10\%$ was used  during the arc melting process. The melting was carried out using low current values for the arc discharge, about 50 Amp.   to minimize any extra loss of bismuth. Samples were remelted several times followed by annealing for 10 days at 900 ${^0}C$.

The X-ray data of the resulting  powders  of polycrystalline sample  was obtained with  a Bruker D8 diffractometer. The unit cell data were derived using DicVol software. The magnetic data reported  here   were measured with  a Quantum Design instrument,  MPMS equipped with a SQUID  (San Diego, CA )  heat capacity more measurements were performed with using  a Physical Property Measurement System PPMS, from Quantum Design.

\section{Results}

\begin{figure}
\includegraphics[width=0.5\textwidth]{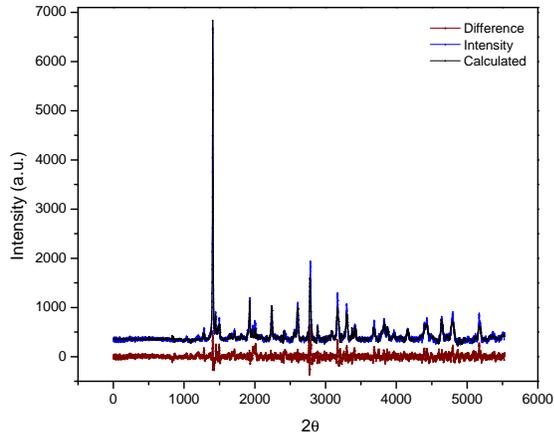}
\caption{(Color on-line) X-ray diffraction pattern and Rietveld refinement, the red  data line shows the difference between the calculated and observed pattern. The space group was orthorhombic $P2$ with lattice parameters;  $a=9.8012(26) \, \AA$, $b=6.0647(6) \, \AA$, $c=6.1663(13) \, \AA$}
\label{fig:XRD}
\end{figure}

At first X-ray powder diffraction analysis Fig.\,\ref{fig:XRD} was used to solve  the crystalline structure and with help of   Rietveld refinement we  carried on and  fit the data yielding the lattice parameters;  $a=9.8012(26) \, \AA$, $b=6.0647(6) \, \AA$, $c=6.1663(13) \, \AA$ in the monoclinic crystalline structure space group type $P2$. Results related to the  atomic coordinates  are shown in Table \ref{table:atomic}. It is important to mention that before trying to solve the crystalline structure, we  tested  various different data bases and found no structure matching it.

\begin{table*}
\caption{ Atomic positions parameters of \ce{HoCuBi} compound. Space group $P2$ (No. 3), $a=9.8012(26) \, \AA$, $b=6.0647(6) \, \AA$, $c=6.1663(13) \, \AA$, $R_{F}=8.7\, \%$}
\begin{center}
\begin{tabular}{cccccc}
\hline\noalign{\smallskip}
Atom & Type of position & x/a & y/b & z/c & Occupation\\ 
\noalign{\smallskip}\hline\noalign{\smallskip}
 Cu & 4e & 0.5244(4)  &  0.5314(5)  &  0.8286(0) & 1.0 \\
 Cu & 4e & 0.6584(4)  &  0.1081(8)  &  0.2697(5) & 1.0 \\
 Bi & 4e & 0.9433(9)  &  0.1374(1)  &  0.2646(7) & 1.0 \\
 Bi & 4e & 0.8390(1)  &  0.5545(1)  &  0.4818(0) & 1.0 \\
 Ho & 4e & 0.8443(5)  &  0.6191(6)  &  0.0314(1) & 1.0 \\
 Ho & 4e & 0.7081(7)  &  0.1662(2)  &  0.7511(3) & 1.0 \\
\noalign{\smallskip}\hline
\end{tabular}
\end{center}
\label{table:atomic}
\end{table*}

\begin{figure}
\includegraphics[width=0.5\textwidth]{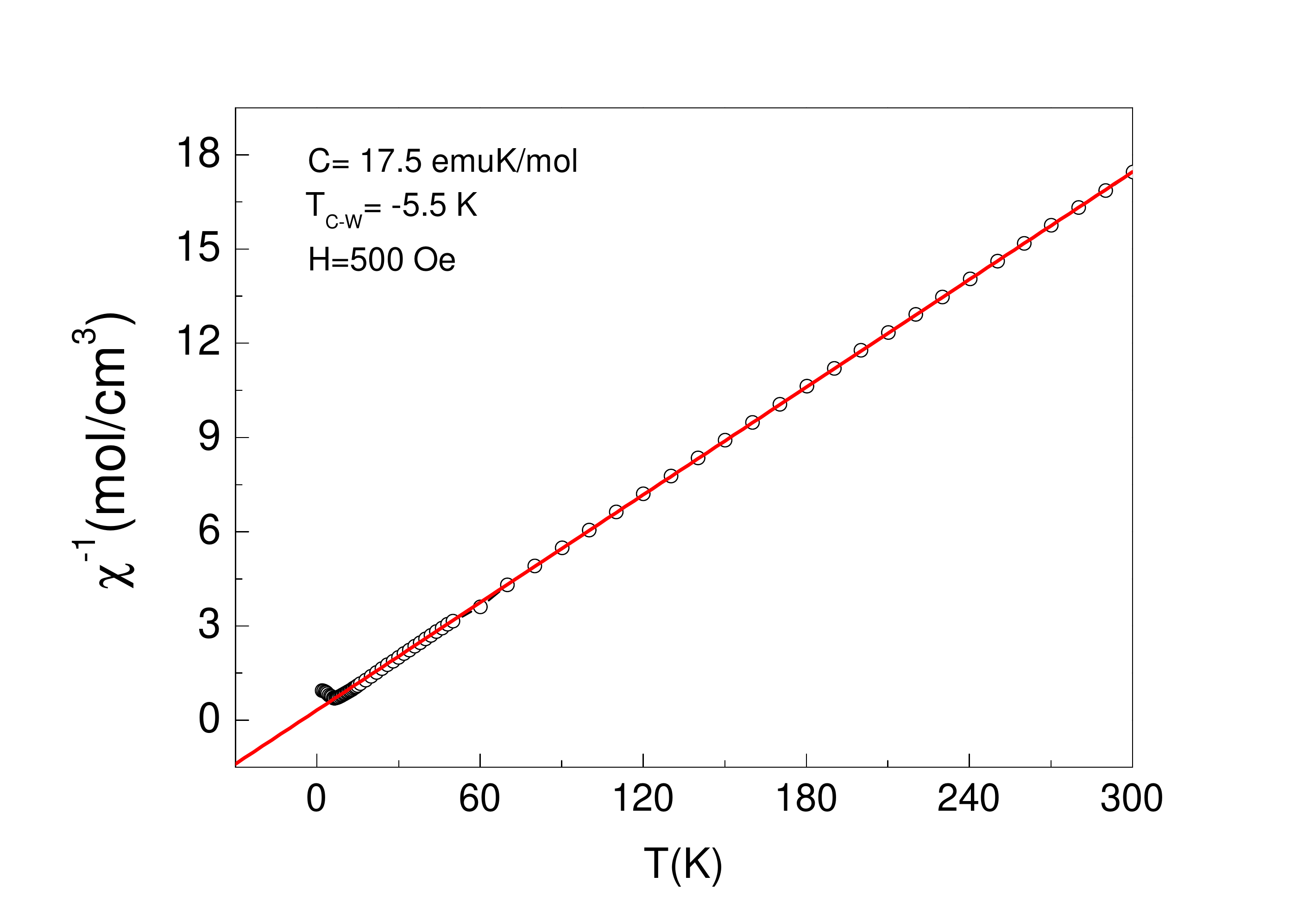}
\caption{(Color on-line) Inverse magnetic susceptibility $\chi^{-1}(T)$.The straight line corresponds to the Curie Weiss law.  At low temperature the data   line presents a change due to the antiferromagnetic transition. The fitting gives a Curie-Weiss  temperature of -5.5 K. The N\`{e}el temperature is $T_{N}=7$ K, and the Curie constant  determined is  $C=17.415 (emuK/mol$)}
\label{fig:MTAjuste}
\end{figure}

The magnetic susceptibility $\chi(T)$ of a specimen of \ce{HoCuBi} shows a sharp peak at low temperature, the  antiferromagnetism   at about  $7$ $K$. This is the  characteristic of a long-range antiferromagnetic order in the sample, with  N\`{e}el temperature  $ T_{N}$ $=$ $7$ $K$. The temperature dependence of the inverse of the magnetic susceptibility and the fit of the Curie-Weiss law to the data are plotted in  Fig.\,\ref{fig:MTAjuste} and confirm the antiferromagnetic transition.\\
\begin{figure}
\includegraphics[width=0.5\textwidth]{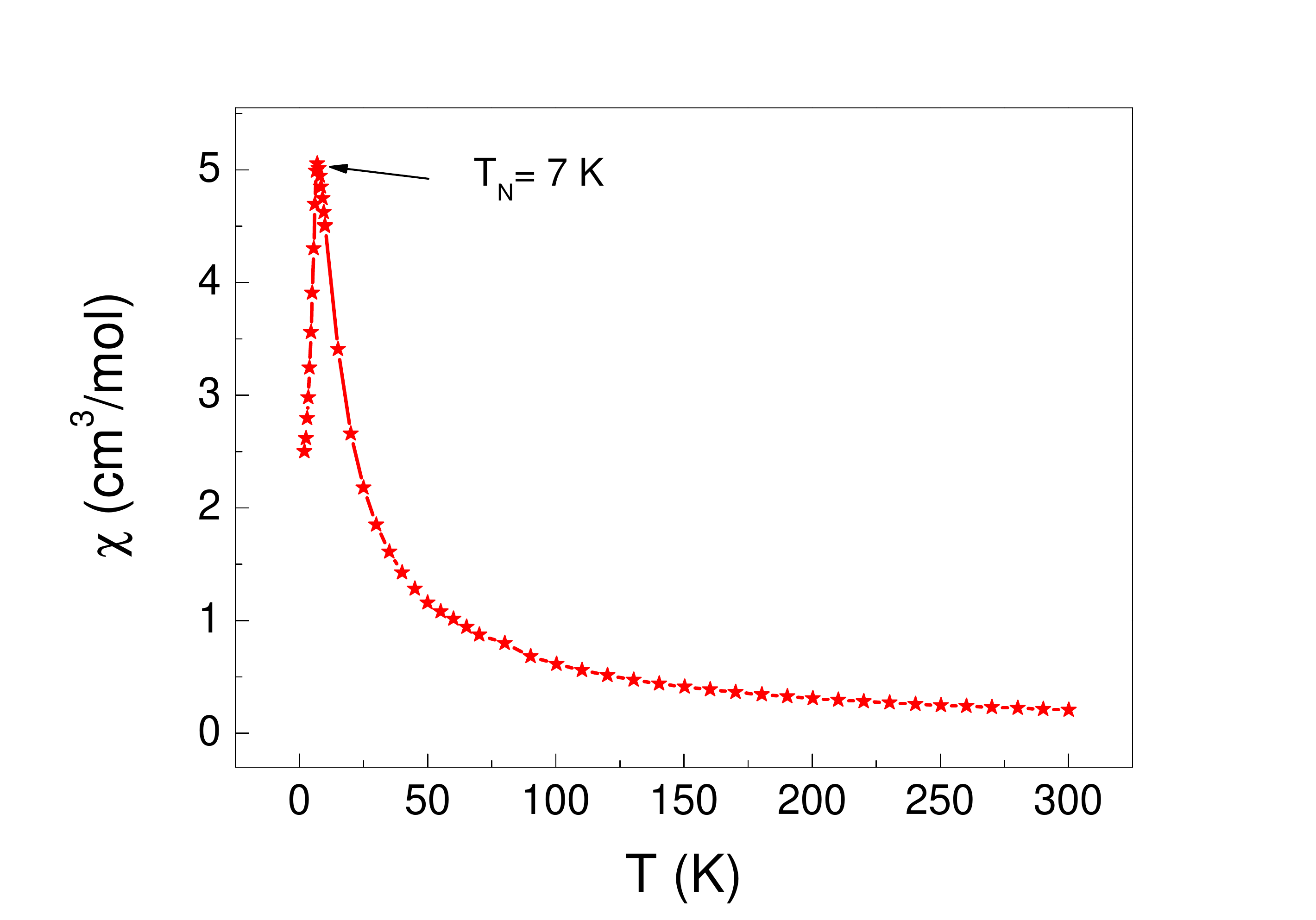}
\caption{(Color on-line) Magnetic susceptibility  $\chi(T)$ from room temperature to 2 K. At low temperatures the peak corresponds to the  maximum of antiferromagnetic ordering ocurring at 7 K, but it is important to remark that the antiferromagnetic transition start at 5.5 K, according to our specific heat measurements.}
\label{fig:MT}
\end{figure}

X-ray diffraction at room temperature shows that this compound crystallizes in a monoclinic $P2$ structure. The temperature dependence of the magnetic susceptibility $\chi$ is shown in Fig.\,\ref{fig:MT}. The magnetic susceptibility clearly indicates the antiferromagnetic ordering at about  $T_{N}=7 K$, showing a steep decrease at $H=1\,T$. The inverse magnetic susceptibility $\chi^{-1}$ is shown in Fig.\,\ref{fig:MTAjuste} which has been adjusted by the Curie-Weiss equation.  The effective magnetic moment is $\mu_{eff}= 11.80$  is close to the theoretical value for holmium $\mu_{eff}= 10.61$ and $\theta_{p}= 88 K$.

As our experimental measurements show at low temperature it was seen an antiferromagnetic transition. In M-H measurements the applied magnetic field can induces  a series of cascade changes from low to high spin flips.  Stryjewski and Giordano, \cite{doi:10.1080/00018737700101433} mentioned that the spin flips may be associated to a first order phase transition, in some cases.

At low temperature below the N\`{e}el temperature we see in M-H measurements an apreciable change of the magnetization with the applied field; this behavior is named in the literature,  as a spin crossover transition. This feature is seen only  at low temperatures indicating a change from low to high spin. This characteristics is observed  in measurements of isothermal M-H  at 2 K and a little above. The corresponding  Brillouin fit indicates variations of $J$ from 1  to 4, and consequently a change of the spins from $s$ = 1/2 to 3/2.
 
\begin{figure}
\includegraphics[width=0.5\textwidth]{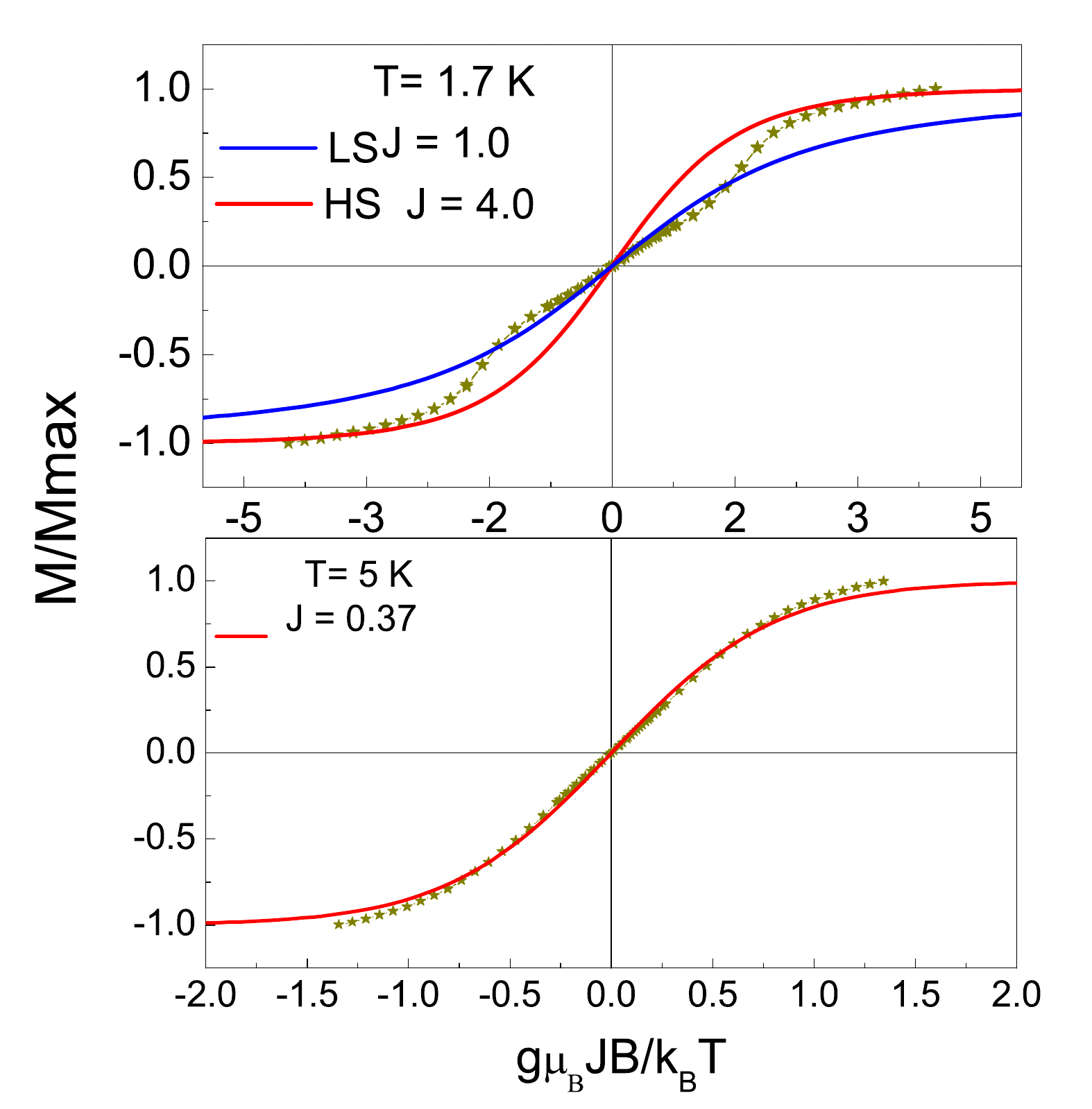}
\caption{(Color on-line) Magnetization normalized versus magnetic field, M/M$_{max}$ vs H. Brillouin  function was used with two distinct $J$ values. The fit  indicates   a change at low applied magnetic field  of a spin crossover from $J$ $1$ to $4$}
\label{fig:Brillouin}
\end{figure}

Specific heat measurement were carried out and the plot $C/T$ versus $T$ is shown in Fig.\,5, applied magnetic field measurements were studied and the plot varying the field from $0$ to $9$ $T$ is shown in Fig.\,6. The specific heat shows the magnetic transition, the sharp peak at $5.5 $ $K$ indicating the antiferromagnetic ordering of this compound. As the magnetic field is increased the transition decreases. Which is the  clear indication that  the antiparallel exchange interaction  of the spins  has been weakened  by the strength of the magnetic field.  

\begin{figure}
\includegraphics[width=0.5\textwidth]{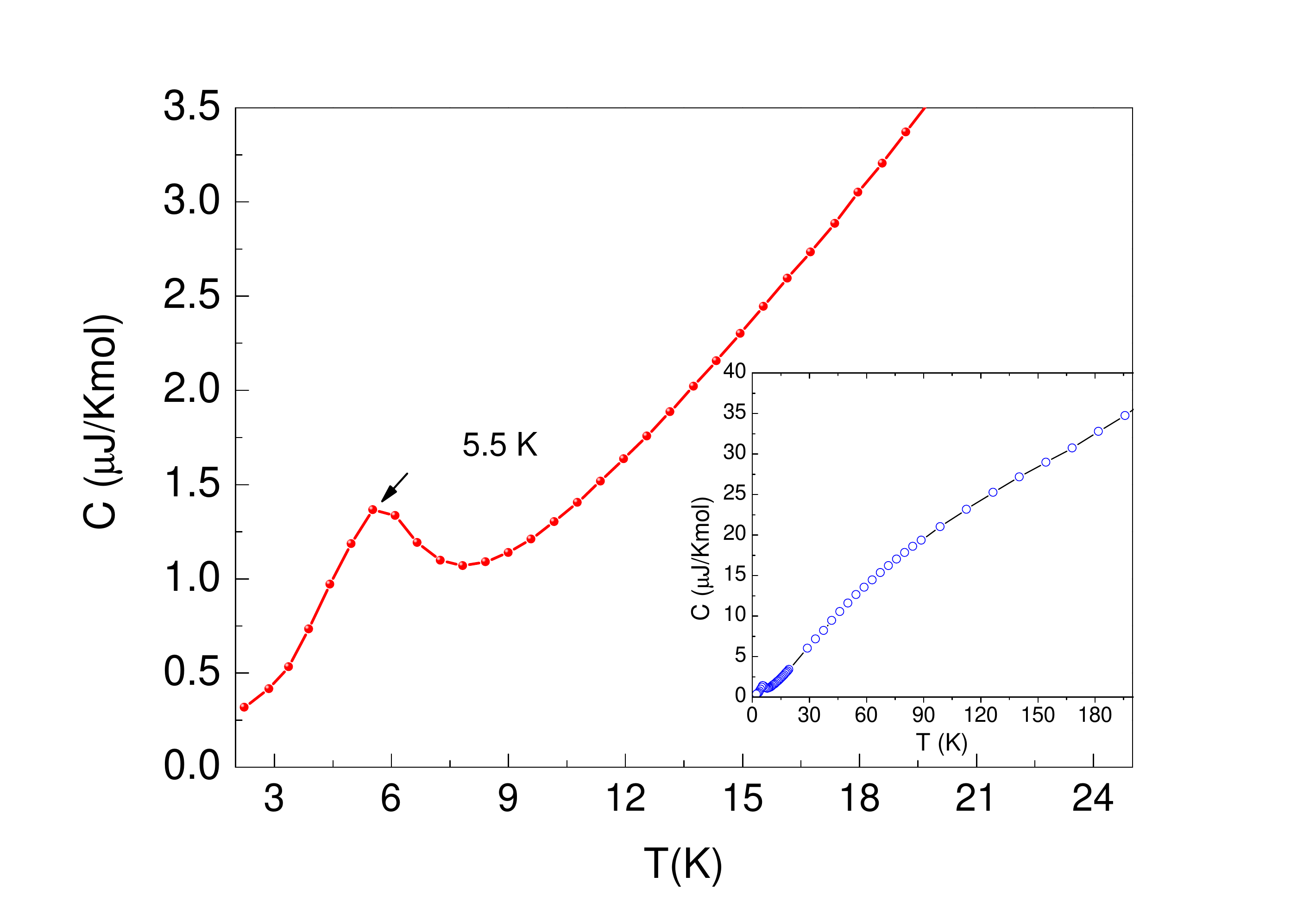}
\caption{(Color on-line) Specific Heat vs Temperature determined at zero applied field. the main panel displays the  maximum of the transition occurring  at 5.5, the onset is  at about 7 K. The insert displays the overall behavior of the specific heat as function of temperature, at low temperature is observed  the   transition}
\label{fig:Cp}
\end{figure}

\begin{figure}
\includegraphics[width=0.5\textwidth]{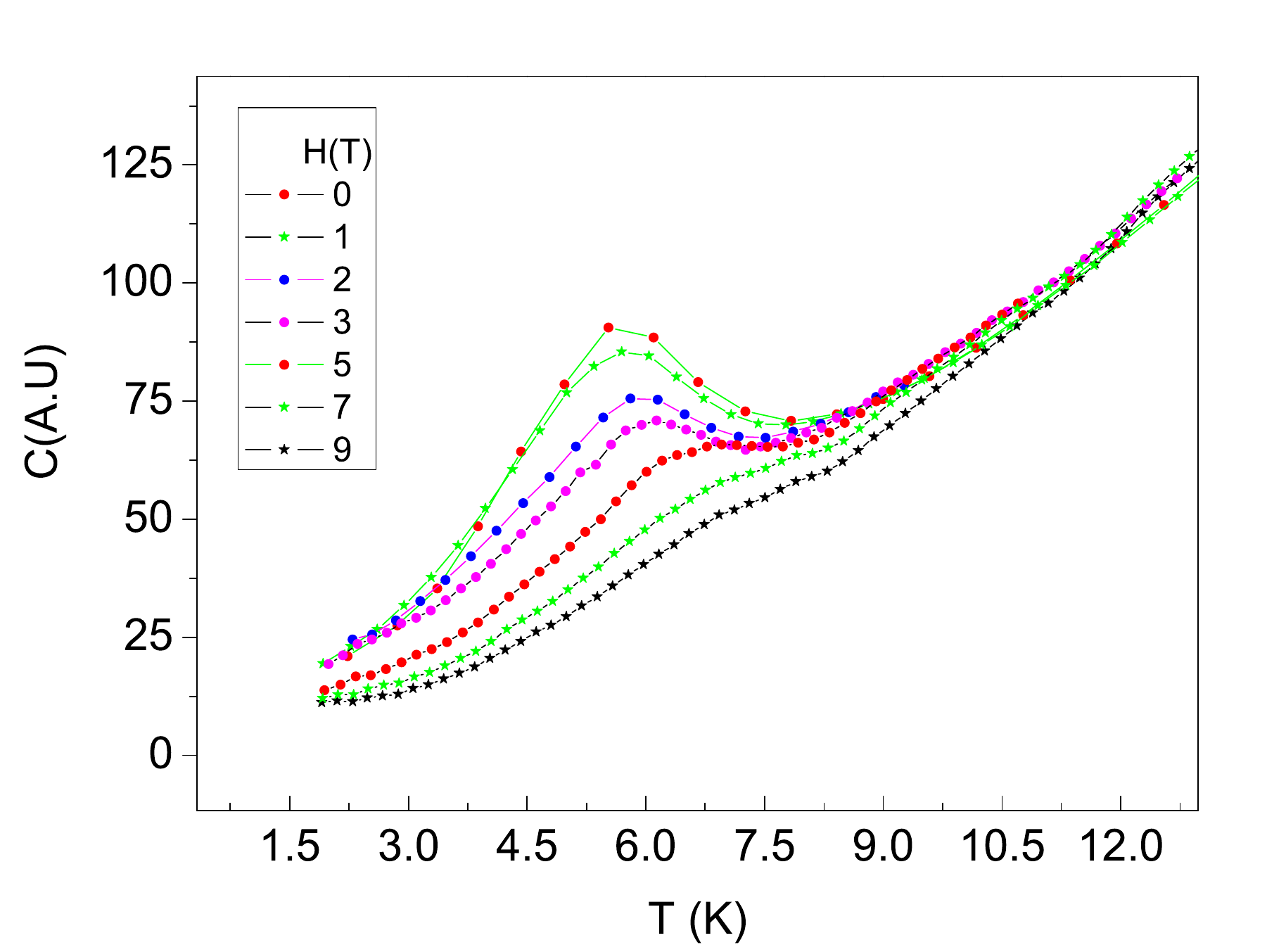}
\caption{(Color on-line) Changes observed in the Specific Heat vs Temperature, as function of magnetic field, from 0 to 9 Tesla, note that the transition decreases as the magnetic field is increased due to the reduced strength of the exchange force. Measurements were vertically displaced by a constant amount in order to have a better resolution.}
\label{fig:Cp1}
\end{figure}

\section{Conclusions}

This investigation shows our recent crystallographic,  magnetic,  and thermal  studies performed on the polycrystalline \ce{HoCuBi} compound which crystallizes in a monoclinic $P2$ structure with lattice parameters;  $a=9.8012(26) \, \AA$, $b=6.0647(6) \, \AA$, $c=6.1663(13) \, \AA$. We found several interesting physical features: an  antiferromagnetic behavior, with  N\`{e}el temperature of $T_{N}=7 \, K$ and effective magnetic moment $\mu_{eff}=11.80$ $\mu_{B}$ which is close to the theoretical value for holmium, a metamagnetic characteristic at low temperature (spin crossover), and influence of the magnetic field in the specific heat transition.

\section*{Acknowledgments}
We thank to  A. Lopez, and A. Pompa-Garcia  del IIM-UNAM), for help in computational and technical problems, also to the grant DGAPA-UNAM  IT100217, and CONACyT for the scholarship to C. Aguilar-Maldonado.

\end{document}